\documentclass[prl,twocolumn,showpacs,preprintnumbers,amsmath,amssymb]{revtex4}

\usepackage{graphicx}
\usepackage{dcolumn}
\usepackage{bm}

\newcommand{\erf}{\mathop{\mbox{erf}}}

\newcommand{\dt}{\mbox{d}t}

\begin{document}

\preprint{}

\title{What is the Time Scale of Random Sequential Adsorption?}

\author{Radek Erban}
 \email{erban@maths.ox.ac.uk}
\author{S. Jonathan Chapman}%
 \email{chapman@maths.ox.ac.uk}
 
\affiliation{%
University of Oxford, Mathematical Institute,
24-29 St. Giles', Oxford, OX1 3LB, United Kingdom
}%

\date{\today}

\begin{abstract}
\noindent
A simple multiscale approach to the diffusion-driven adsorption 
from a solution to a solid surface is presented. The model
combines two important features of the adsorption process:
(i) the kinetics of the chemical reaction between adsorbing
molecules and the surface; and (ii) geometrical constraints on
the surface made by molecules which are already adsorbed.
The process (i) is modelled in a diffusion-driven context,
i.e. the conditional probability of adsorbing a molecule 
provided that the molecule hits the surface is related to the 
macroscopic surface reaction rate. The geometrical constraint (ii) 
is modelled using random sequential adsorption (RSA), which
is the sequential addition of molecules at random positions on a 
surface; one attempt to attach a molecule is made per one RSA
simulation time step. By coupling RSA with the diffusion of molecules 
in the solution above the surface the RSA simulation time step is 
related to the real physical time. The method is illustrated 
on a model of chemisorption of reactive polymers to a virus surface.
\end{abstract}

\pacs{68.43.-h, 87.15.Rn}
\maketitle

Random sequential adsorption (RSA) is a classical model of  
irreversible adsorption (e.g. chemisorption) 
\cite{Evans:1993:RCS}. 
Given a sequence of times $t_k$, $k=1, 2, 3, \dots$, an attempt
is made to attach one object (e.g. a molecule) to the surface at 
each time point $t=t_k$. If the attempt is successful (i.e. 
if there is enough space on the surface to place the molecule),
the object is irreversibly adsorbed. It cannot further
move or leave the structure and it covers part of the surface,
preventing other objects from adsorbing in its neighbourhood (e.g.
by steric shielding in the molecular context).

In the simplest form, RSA processes are formulated as attempting 
to place one object per RSA time step, expressing the 
simulation time in units equal to the number of RSA 
time steps $k$ rather than in real physical time $t_k$.
Such an approach is useful to compute the maximal
(jamming) coverage of the surface. To apply RSA models
to dynamical problems, it is necessary to relate the time
of the RSA simulation $k$ and the real time $t_k$. 
This is a goal of this paper. We consider that the adsorbing 
objects are molecules which can covalently attach to 
the binding sites on the surface. We couple the RSA model
with processes in the solution above the surface to study the 
irreversible adsorption of molecules in real time. The time 
$(t_{k} - t_{k-1})$ between the subsequent attempts to place 
a molecule is in general a non-constant 
function of $k$ which depends on the kinetics of the chemical 
reaction between the adsorbing molecules and 
the surface, and on the stochastic reaction-diffusion 
processes in the solution above the surface. We illustrate
our method with an example of the chemisorption of reactive
polymers to a virus surface \cite{Erban:2006:DPI,Erban:2006:CPS}.
Finally, we show that the stochastic simulation in the solution 
can be substituted by a suitable deterministic partial differential 
equation which decreases the computational intensity of the 
algorithm. We show that it is possible to get the values 
of $t_k$ without doing extensive additional stochastic simulations.

\begin{figure}[ht]
\includegraphics[width=70mm]{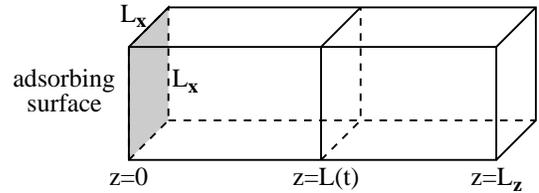}
\vskip -3mm
\caption{{\it Three-dimensional cuboid domain.}}
\label{figdomain}
\end{figure}

We consider a three-dimensional cuboid domain $L_x \times L_x \times L_z$ 
in which molecules diffuse (see Fig.\,\ref{figdomain}). The side $z=0$ 
of area $L_x \times L_x$ is assumed to be adsorbing, i.e. containing 
binding sites to which molecules can covalently attach. Our goal is 
to couple  RSA on the side $z=0$ with stochastic 
reaction-diffusion processes in the solution above the adsorbing
surface. 
Since those molecules which are far from the surface will have
little influence on the adsorbtion process, it is a waste of resources
to compute their trajectories. We will therefore
carefully truncate
our computational domain to that which is effectively influenced by
the reactive boundary at $z=0$, which we denote by $z<L(t)$. Note that
$L(t)$ is not fixed but a function of time---the formula for it will
be derived later. 
Suppose that there are $N(t)$ diffusing molecules 
in the cuboid domain $L_x \times L_x \times L(t)$. Let us 
denote the $z$-coordinate of the $i$-th molecule by 
$z_i(t)$, treating molecules as points in the solution
in what follows. Choosing a time step $\Delta t$, 
we compute $z_i(t+\Delta t)$ from $z_i(t),$
$i = 1, \dots, N(t),$ by  
\begin{equation}
z_i(t + \Delta t) = z_i(t) + \sqrt{2 D_i \, \Delta t} \; \xi_i,
\label{discretestochevol}
\end{equation}
where $\xi_i$ is a normally distributed random variable with zero
mean and unit variance and $D_i$ is the diffusion constant
of the $i$-th molecule. 
In principle, we should model the behaviour of molecules as 
three dimensional random walks in the cuboid domain 
$L_x \times L_x \times L(t)$, i.e. there should be equations 
analogous to (\ref{discretestochevol}) for the $x$ and 
$y$ coordinates too. However, we can often assume that 
$L(t) \gg L_x$ in applications. Choosing the time step 
$\Delta t$ large enough that a molecule travels over 
distances comparable to $L_x$ during one time step,
we can assume that the molecules are effectively well-mixed 
in the $x$ and $y$ directions on this time scale. 
Consequently, the $x$ and $y$ coordinates of molecules 
do not have to be simulated. If the original adsorbing
surface is large, one often models by RSA only a representative
part of it, i.e. a square $L_x \times L_x$ which
contains a relatively large number of binding sites, but still
satisfies
$L_x \ll L(t)$. The 
diffusion of molecules (\ref{discretestochevol}) 
is coupled with other processes in the solution and 
on the surface as follows.

\noindent
{\bf Chemical reactions in the solution}:
Our illustrative example is the polymer coating of viruses 
\cite{Erban:2006:DPI,Erban:2006:CPS}. In this case, the polymer molecules 
have reactive groups which can covalently bind to the surface. The
reactive groups also hydrolyse in solution. Assuming that there is one
reactive group per polymer molecule (such a polymer is called
semitelechelic), we have effectively one chemical reaction
in the solution - a removal of the reactive polymers from the solution
with rate $\lambda$ \cite{Subr:2006:CDC}. Assuming 
that $\lambda \Delta t \ll 1$,
the stochastic modelling of the process in the solution
is straightforward. At each time step, the $i$-th molecule
moves according to (\ref{discretestochevol}). We then generate
a random number $r_i$ uniformly distributed on the interval $[0,1]$.
If $r_i < \lambda \Delta t$, we remove the molecule from the system.
More complicated reaction mechanisms in the solution can be treated
using stochastic simulation algorithms which have been
proposed for reaction-diffusion processes in the literature 
\cite{Andrews:2004:SSC,Hattne:2005:SRD,Isaacson:2006:IDC}.
In our case, we treat diffusion using the discretized version of 
Smoluchowski equation (\ref{discretestochevol}). Consequently, we can
follow Andrews and Bray \cite{Andrews:2004:SSC} to 
introduce higher-order reactions to the system. 

\noindent
{\bf Adsorption to the surface}: The surface $L_x \times L_x$ at $z=0$ 
is assumed to be adsorbing. We use a simple version of the RSA model from 
\cite{Erban:2006:CPS} which postulates that the binding sites
on the surface lie on a rectangular lattice. Binding a polymer
to a lattice site prevents the binding of another polymer
to the neighbouring lattice sites through steric 
shielding, i.e. we consider RSA with the nearest neighbour exclusion
as a toy model of adsorption \cite{Evans:1993:RCS}.
Such a RSA model can be simulated on its own as shown 
in \cite{Erban:2006:CPS}. In this paper, we simulate it
together with the $z$-variables of molecules in the 
solution (\ref{discretestochevol}) to get the RSA
evolution in real physical time. Whenever a molecule hits 
the boundary $z=0$, it is adsorbed with some probability, 
and reflected otherwise. This partially adsorbing boundary 
condition is implemented in the RSA context as follows: 

\smallskip
\leftskip 2mm
\rightskip 2mm
 
\noindent 
{\bf (a)} If $z_i(t+\Delta t)$ computed by (\ref{discretestochevol}) 
is negative then, with probability $P \sqrt{\Delta t}$, 
we attempt one step of the RSA algorithm with the $i$-th molecule.
If the $i$-th molecule is adsorbed, we remove it from the solution.
Otherwise, we put $z_i(t+\Delta t) = - z_i(t) - \sqrt{2 D_i \, \Delta t} 
\; \xi_i$.

\smallskip

\noindent 
{\bf (b)} If $z_i(t+\Delta t)$ computed by  (\ref{discretestochevol}) is 
positive then, with probability 
$\exp[- x_i(t) x_i(t + \Delta t)/(D \Delta t)] P \sqrt{\Delta t}$, 
we attempt one step of the RSA algorithm with the $i$-th molecule.
If the $i$-th molecule is adsorbed, we remove it from the solution.

\smallskip
\leftskip 0mm
\rightskip 0mm

\noindent
Here, $P$ is a positive constant which can be related to the rate 
constant of the chemical reaction between the binding sites on the virus 
surface and the reactive groups on the polymer \cite{Erban:2006:RBC}.
This relation depends on the stochastic model of diffusion
and for  equation (\ref{discretestochevol}) is given 
later -- see formula (\ref{RobinBD}). 
Conditions (a)--(b) state that only the fraction $P \sqrt{\Delta t}$
of molecules which hit the boundary have a chance to create a
chemical bond (provided that there is no steric shielding).
Obviously, if $z_i(t + \Delta t)$ computed by 
(\ref{discretestochevol}) is negative, a molecule has hit the 
boundary. This case is incorporated in (a). However, Andrews and 
Bray \cite{Andrews:2004:SSC} 
argue that there is a chance that a molecule hit the boundary 
during the finite time step $\Delta t$ even if $z_i(t + \Delta t)$ 
computed by (\ref{discretestochevol}) is positive; that is,
during the time interval $[t,t+\Delta t]$ the molecule might
have crossed to $z_i$ negative and then crossed back to 
$z_i$ positive again. They found
that the probability that the molecule hit the boundary
$z=0$ at least once during the time step  $\Delta t$
is  $\exp[- z_i(t) z_i(t + \Delta t)/(D \Delta t)]$
for $z_i(t) \ge 0$, $z_i(t + \Delta t) \ge 0$. This formula
is used in (b).

\begin{figure*}
\centerline{
\includegraphics[height=2.625in]{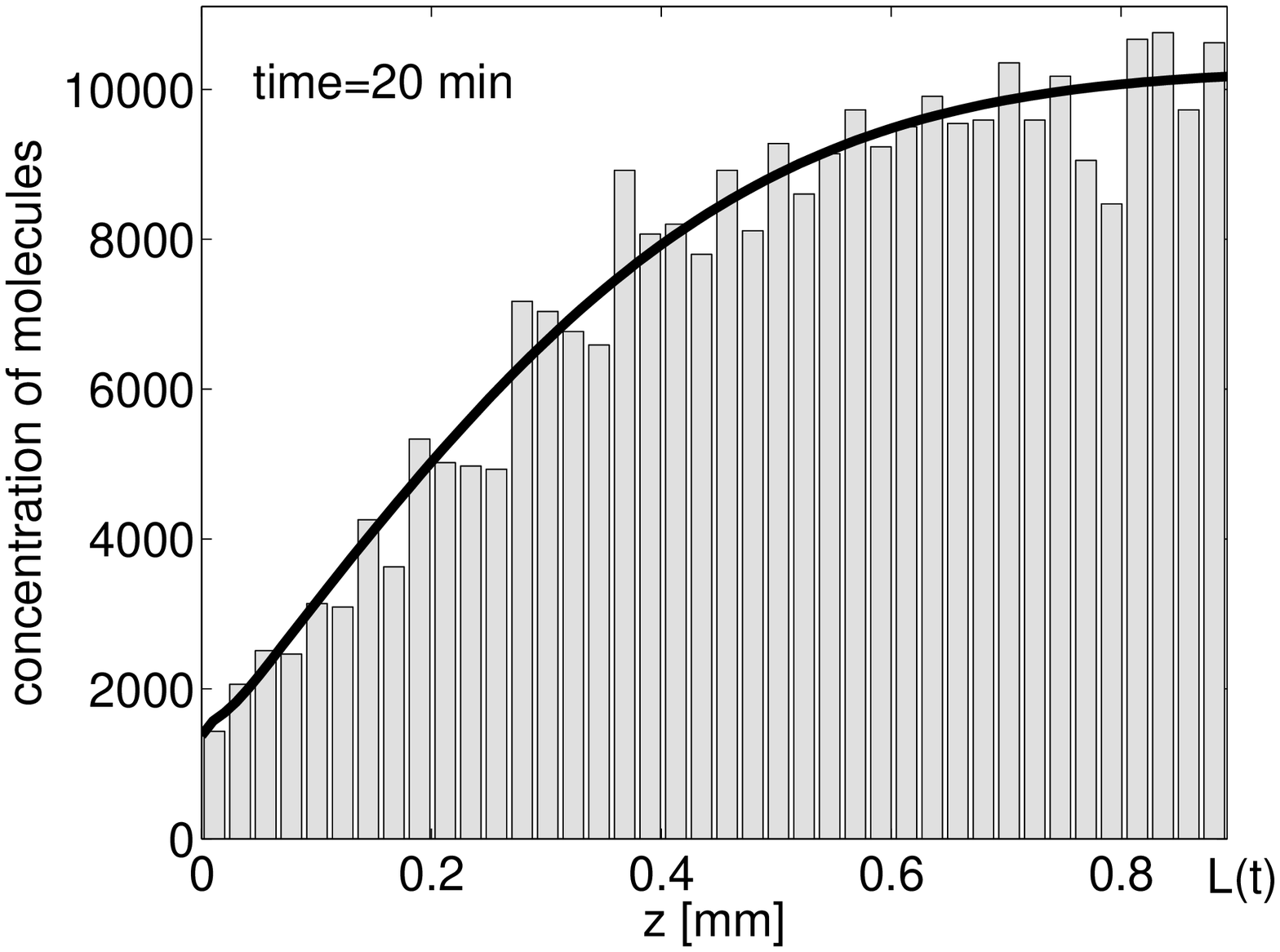}
\!\!\!\!\!\!
\includegraphics[height=2.625in]{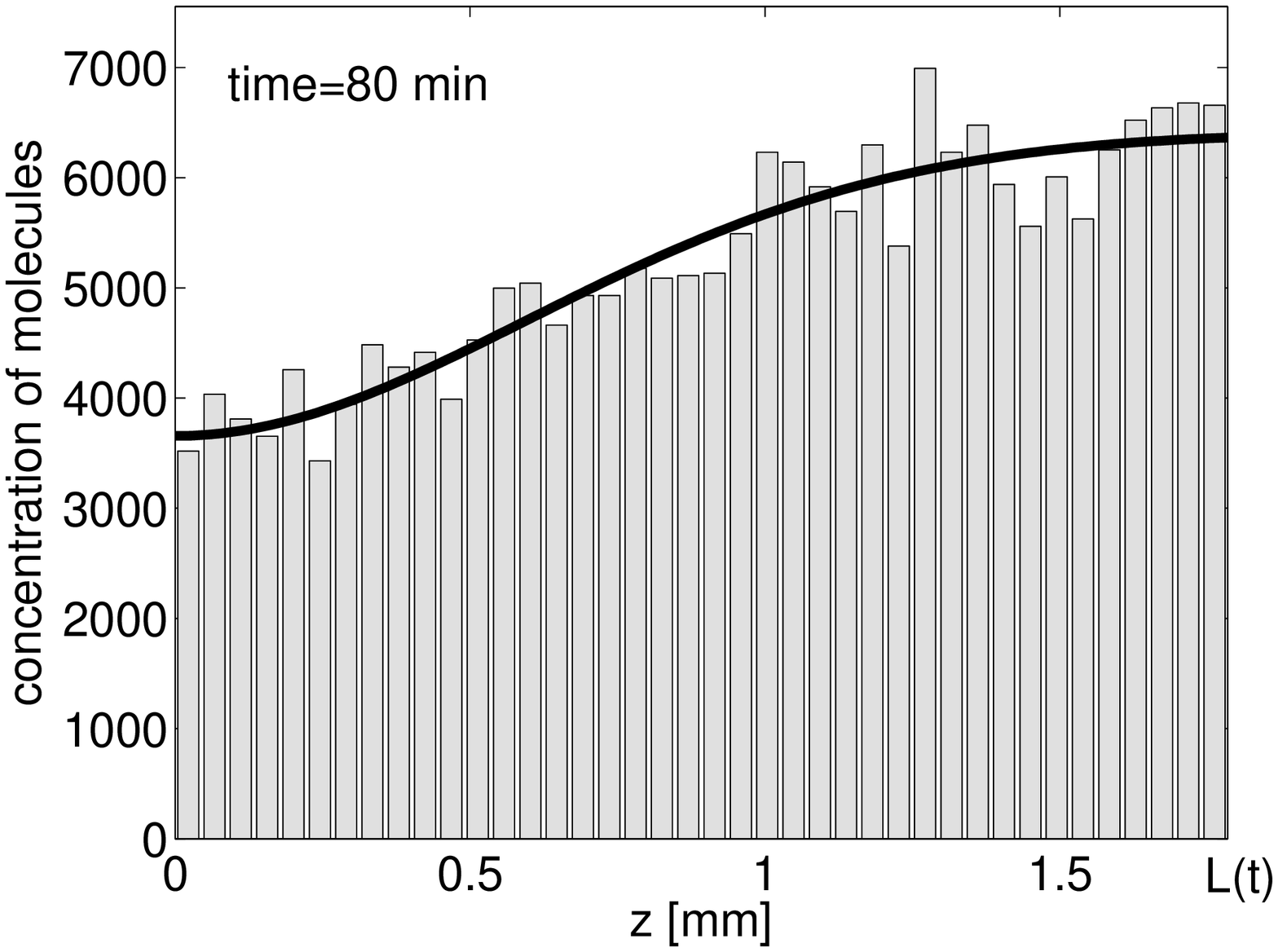}
}
\caption{{\it Concentration of polymer molecules in the solution 
above the adsorbing surface $z=0$ at times 20 and 80 minutes.}}
\label{figRSAcoupling}
\end{figure*}

\noindent
{\bf Numerical results}: 
It is important to note that the boundary conditions 
(a)--(b) can be used for any RSA algorithm and for 
any set of reactions in the solution. To apply
it to the virus coating problem, we have to specify
some details of the model. First of all, it can be estimated
that the average distance between the binding sites is about
1 nm \cite{Erban:2006:DPI}. We choose $L_x = 100$ nm.
Therefore, there are about $10,000$ binding sites on the 
adsorbing side $z=0$. We use RSA on a $100 \times 100$ lattice
with the nearest neighbour exclusion, which is a special case
of the model from \cite{Erban:2006:CPS}.
We consider a monodisperse solution of semitelechelic 
50 kDa polymers, i.e. $D_i \equiv D$ where 
$D=5 \times 10^{-5} \; \mbox{mm}^2 \; \mbox{s}^{-1}$
\cite{Prokopova:2001:PPD}.   
The rate of hydrolysis of the reactive groups on polymers can 
be estimated from data in \cite{Subr:2006:CDC} as 
$\lambda = 1.3 \times 10^{-4} \, \mbox{s}^{-1}.$
We choose $P = 1 \; \mbox{s}^{-1/2}$.
Since we simulate the behaviour
of polymer molecules in solution only along the $z$-direction,
we express the concentration of polymer $c(z,t)$ in numbers of polymer
molecules per volume $L_x \times L_x \times [1 \; \mbox{mm}]$
where $L_x = 10^{-4}$ mm is fixed. A typical experiment starts
with a uniform concentration of reactive polymers.
Considering that the initial concentration of 50 kDa polymer
is 0.1 g/l, we obtain the initial condition $c_0= 1.2 \times 10^{4}$ 
molecules per mm of the height above the surface (let us note 
that the units of the ``one-dimensional" concentration $c(z,t)$ 
are molecules/mm because $L_x$ is considered fixed).
Next, we have to specify $L(t)$ (see Fig.\,\ref{figdomain}), 
i.e. we want to find the region of the space which is effectively 
influenced by the boundary condition at $z=0$. To that end, we note
that the concentration $c(z,t)$ satisfies the partial differential
equation
\begin{equation}
\frac{\partial c}{\partial t} =
D \frac{\partial^2 c}{\partial z^2}   - \lambda c.
\label{parab}
\end{equation}
Now any partially reacting boundary will have
less impact on the spatial profile of $c(z,t)$ than  
perfect adsorption at $z=0$. 
Thus we may find an upper bound for the region of influence of the
boundary by solving (\ref{parab}) subject to 
\begin{equation}
c(0,t) = 0,
\qquad
\lim_{z \to \infty} c(z,t) = c_0 \exp [-\lambda t],
\label{boundconditions}
\end{equation}
for $t \in [0,\infty)$, and the initial condition 
$c(z,0) = c_0,$ for $z \in [0,\infty)$. 
The solution of (\ref{parab})--(\ref{boundconditions}) is
\begin{equation}
c(z,t) = c_0 \exp [-\lambda t] \erf \left( \frac{z}{2 \sqrt{D t}} \right)
\label{solution}
\end{equation}
where $\erf(\cdot)$ denotes the error function.
Defining $\omega = \mbox{erf}^{-1}(0.99) \doteq 1.821$ we set
\begin{equation}
L(t) \equiv 2 \omega \sqrt{D t}.
\label{defLz}
\end{equation}
Then $c(L(t),t) = 0.99 \, c_0 \exp [-\lambda t]$, so that 
the concentration of the reactive polymer at point 
$z=L(t)$ 
at time $t$ is equal to 99 \% of the polymer concentration at points
which are ``infinitely far" from the adsorbing boundary. In particular,
we can assume that the adsorbing boundary effectively influences the polymer
concentration only at  heights $z \in [0,L(t)]$ above the boundary
and we can approximate $c(z,t) \sim c_0 \exp [-\lambda t]$ for $z > L(t).$
Formula (\ref{defLz}) specifies the computational domain 
as $L_x \times L_x \times L(t)$ at each time 
(see Fig.\,\ref{figdomain}).

The results of the stochastic simulation of the solution
above the surface are shown in Fig.\,\ref{figRSAcoupling} 
as grey histograms.
To simulate the behaviour of $N(t)$ reactive polymers, we consider 
only their $z$-positions. We use $\Delta t = 10^{-2}$ s and
we update the $z$-positions of molecules during one time step according to
(\ref{discretestochevol}). At each time step, we also generate
a uniformly distributed random number $r_i$ and we remove the
$i$-th molecule from the system if $r_i < \lambda \, \Delta t$. 
We work in the one-dimensional domain $[0,L(t)]$ where $L(t)$ is given
by (\ref{defLz}). The RSA boundary condition
at $z=0$ is implemented using (a)--(b) described 
above. The right boundary increases 
during one time step by $\Delta L(t) = L(t + \Delta t) - L(t)$.
During each time step, we have to put on average 
$m(t)=c_0 \exp [-\lambda t] \Delta L(t)$ molecules into the interval 
$[L(t), L(t + \Delta t)]$. This is done
as follows. We put $\lfloor m(t) \rfloor$ 
molecules at random positions in the interval
$[L(t), L(t + \Delta t)]$, where $\lfloor \cdot \rfloor$
denotes the integer part. Moreover, we generate
random number $r_{\Delta t}$ uniformly distributed in $[0,1]$
and we add one molecule at a random position in
the interval $[L(t), L(t + \Delta t)]$ if
$r_{\Delta t} < m(t) - \lfloor m(t) \rfloor.$
This will ensure that we put on average $m(t)$ molecules to the 
interval $[L(t), L(t + \Delta t)]$ during one time step.

Introducing the moving boundary decreases the computational
intensity of the model. Initially
we simulate a relatively small region with a high concentration 
of reactive polymers. The simulated region increases with time but
the concentration of reactive molecules decreases with rate 
$\lambda$. Using (\ref{solution}), it can be computed that the 
maximal number of simulated polymers in solution is
achieved at time $t_m = (2 \lambda)^{-1} \doteq 64$
min (and is about $8 \times 10^3$ molecules for our parameter
values).

\begin{figure}[t]
\centerline{\;\;
\includegraphics[height=2.625in]{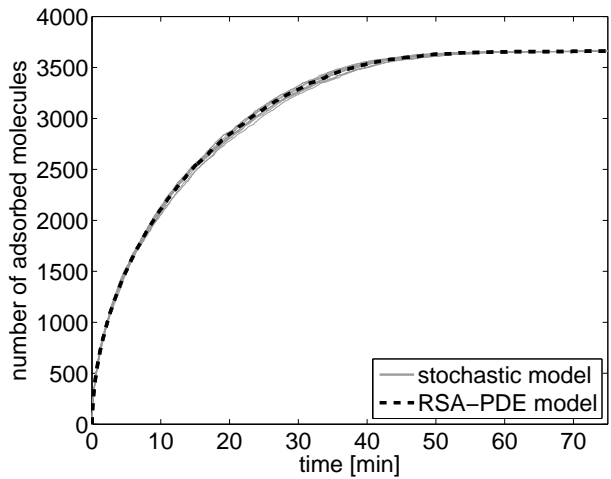}
}
\caption{{\it Number of polymer molecules adsorbed to the RSA lattice
as a function of the real physical time.}}
\label{fignoadsorbed}
\end{figure}

The number of polymers adsorbed to the RSA surface at $z=0$
as a function of real physical time is shown 
in Fig.\,\ref{fignoadsorbed}.
Since the polymer solution is assumed to be monodisperse, we 
can run the RSA algorithm first and record the times 
$k_1,$ $k_2$, $k_3$, \dots (expressed in numbers of the RSA 
time steps) of successful attempts to place the polymer 
on the RSA lattice. Then the stochastic simulation of
the reaction-diffusion processes in the solution can use
$k_1,$ $k_2$, $k_3$, \dots as its input. We will shortly
consider another approach to the problem, replacing the 
stochastic simulation of the solution by the continuum
limit (\ref{parab}) with a suitable Robin boundary condition.
To enable a direct comparison of the two approaches, we 
use the same sequence 
$k_1,$ $k_2$, $k_3$, \dots in ten realizations of the full stochastic model of 
adsorption; the results are shown as grey solid
lines in  Fig.\,\ref{fignoadsorbed}. 

\smallskip
\noindent
{\bf RSA-PDE approach}: Moving the right boundary $L(t)$ is one way to 
decrease the computational intensity of the problem. Another possibility
is to use the deterministic equation (\ref{parab}) together
with a Robin boundary condition 
\begin{equation}
\frac{\partial c}{\partial z} (0,t)
=
\alpha_{RSA}(t) \frac{2 P}{\sqrt{D \pi}} \,\, c(0,t)
\label{RobinBD}
\end{equation}
at the adsorbing boundary $z=0$. Here, the fraction 
$2 P /\sqrt{D\pi}$ corresponds to the rate of the chemical reaction
between the adsorbing boundary and the diffusing 
molecules in the solution -- see \cite{Erban:2006:RBC}
for the derivation of this formula and further discussion. 
Factor $\alpha_{RSA}(t) \in \{0,1\}$ provides the
coupling between the RSA model and (\ref{parab}).
To find the value of $\alpha_{RSA}(t)$, we 
estimate the number of attempts to place the polymer 
on the RSA lattice by
$$
\kappa(t) 
=
\left\lfloor 
\int_0^t \frac{2 P \sqrt{D}}{\sqrt{\pi}} c(0,t) \dt
\right\rfloor
$$
where $\lfloor \cdot \rfloor$ denotes the integer part
\cite{Erban:2006:RBC}.
We start with $\alpha_{RSA}(t)=1$ and we solve (\ref{parab}) 
and (\ref{RobinBD}) numerically. Whenever $\kappa(t)$ increases
by 1, we attempt one step of the RSA.
If the attempt is successful, we put $\alpha_{RSA}(t)=1$.
If the attempt to place the molecule is not successful,
we put $\alpha_{RSA}(t)=0$. Thus 
$\alpha_{RSA}(t)$ has only two values, 0 and 1, and changes
at computed time points depending on the output of the
RSA simulation. 
We call this procedure the RSA-PDE approach. It also leads to the sequence
of real physical times $t_{k_1}$, $t_{k_2}$, $t_{k_3}$  $\dots$, 
of successful attempts to place the polymer on the RSA lattice.
The numerical solution of equation (\ref{parab}) with the 
Robin boundary condition (\ref{RobinBD}) at $z=0$ is presented 
in Fig.\,\ref{figRSAcoupling} as the solid line for
comparison. We also plot the number of adsorbed polymers
as a function of the real time as the dashed line in 
Fig.\,\ref{fignoadsorbed}. To enable the direct comparison,
we run the RSA algorithm first and we record the times 
of successful attempts to place the polymer on the lattice. 
We obtain the sequence $k_1,$ $k_2$, $k_3$, \dots of times 
expressed in number of RSA time steps. This sequence is used
in both the stochastic model (10 realizations plotted in
Fig.\,\ref{fignoadsorbed} as grey solid lines) and the
RSA-PDE approach
(dashed line in Fig.\,\ref{fignoadsorbed}). The comparison
of the results obtained by the full stochastic model and 
by the RSA-PDE model is 
excellent. 

\noindent
{\bf Conclusion}: 
We have presented a method to perform RSA simulation
in real physical time. The key part of the method is the
boundary condition (a)--(b) which can be coupled with any
reaction-diffusion model in the solution and any RSA algorithm.
We illustrated this fact on a simple model of the polymer coating
of viruses. Moreover, we showed that the RSA algorithm
can be coupled with (\ref{parab}) using the Robin boundary
condition (\ref{RobinBD}) to get  comparable results. The Robin 
boundary condition (\ref{RobinBD}) is also not restricted
to our illustrative example. It can be used for the coupling
of any RSA model with the PDE model of the
reaction-diffusion processes in the solution above the
adsorbing surface.

\noindent
{\bf Acknowledgments}: 
This work was supported by the Biotechnology and Biological 
Sciences Research Council.

\end{document}